\documentclass[9pt,twocolumn,twoside]{pnas-new}
\usepackage{acronym}

\templatetype{pnasresearcharticle}

\title{Visualizing Conical Intersection Passages via Vibronic Coherence Maps Generated by Stimulated Ultrafast X--Ray Raman Signals}

\author[a]{Daniel Keefer}
\author[b]{Thomas Schnappinger} 
\author[b]{Regina de Vivie-Riedle}
\author[a,1]{Shaul Mukamel}

\affil[a]{Departments of Chemistry and Physics and Astronomy, University of California, Irvine, California 92697-2025, USA}
\affil[b]{Department Chemie, Ludwig-Maximilians-Universität München, D-81377 München, Germany}

\leadauthor{Keefer} 

\significancestatement{The capabilities of modern x--ray sources have substantially increased the temporal and spectral resolutions for the observation of elementary molecular events. Theoretically, it has been demonstrated that these unique features can be exploited to probe vibronic coherences that emerge when molecules pass through conical intersections. Here, we demonstrate this technique for the photorelaxation of the RNA-nucleobase uracil after UV excitation. The generated coherence maps reveal the path of the nuclear wavepacket on the excited state surface in the vicinity of the conical intersection. This information is directly accessible from the spectroscopically recorded signal. Being transferrable to other molecules, these coherence maps will provide detailed visualizations of non-adiabatic passages in the future.}

\authorcontributions{D.K. and S.M. designed the project. D.K. performed quantum dynamics, and calculated the polarizability, signals and Wigner spectrograms. T.S. computed the non-adiabatic couplings. D.K. and S.M. analyzed and interpreted the results, and wrote the manuscript with critical input from T.S. and R.d-V.R. All authors were involved in the scientific discussion of the results and the paper.}
\authordeclaration{The authors declare no conflict of interest.}

\correspondingauthor{\textsuperscript{1}To whom correspondence should be addressed. E-mail: smukamel@uci.edu}


\begin{abstract}
	The rates and outcomes of virtually all photophysical and photochemical processes are determined by Conical Intersections. These are regions of degeneracy between electronic states on the nuclear landscape of molecules where electrons and nuclei evolve on comparable timescales and become strongly coupled, enabling radiationless relaxation channels upon optical excitation. Due to their ultrafast nature and vast complexity, monitoring Conical Intersections experimentally is an open challenge. We present a simulation study on the ultrafast photorelaxation of uracil, which demonstrates a new window into Conical Intersections obtained by recording the transient wavepacket coherence during this passage with an x--ray free electron laser pulse. We report two major findings. First, we find that the vibronic coherence at the conical intersection lives for several hundred femtoseconds and can be measured during this entire time. Second, the time-dependent energy splitting landscape of the participating vibrational and electronic states is directly extracted from Wigner spectrograms of the signal. These offer a novel physical picture of the quantum Conical Intersection pathways through visualizing their transient vibronic coherence distributions. The path of a nuclear wavepacket around the Conical Intersection is directly mapped by the proposed experiment.
\end{abstract}

\dates{This manuscript was compiled on \today}
\doi{\url{www.pnas.org/cgi/doi/10.1073/pnas.XXXXXXXXXX}}

\begin{document}
	
	\acrodef{coin}[CoIn]{Conical intersection}
	\acrodef{fc}[FC]{Franck-Condon}
	\acrodef{fwhm}[FWHM]{full width at half maximum}
	\acrodef{nac}[NAC]{non-adiabatic coupling}
	\acrodef{oct}[OCT]{optimal control theory}
	\acrodef{pes}[PES]{potential energy surface}
	\acrodef{qd}[QD]{quantum dynamics}
	\acrodef{si}[SI]{Supporting Information}
	\acrodef{tdse}[TDSE]{time dependent Schr\"odinger equation}
	\acrodef{wf}[WF]{wavefunction}
	\acrodef{wp}[WP]{wavepacket}
	
	\newcommand*{\szero}{S\textsubscript{0} }
	\newcommand*{\sone}{S\textsubscript{1} }
	\newcommand*{\stwo}{S\textsubscript{2} }
	\newcommand*{\stwoone}{S\textsubscript{2}/S\textsubscript{1} }
	\newcommand*{\tc}{TRUECARS }
	
	\maketitle
	\thispagestyle{firststyle}
	\ifthenelse{\boolean{shortarticle}}{\ifthenelse{\boolean{singlecolumn}}{\abscontentformatted}{\abscontent}}{}
	
	% If your first paragraph (i.e. with the \dropcap) contains a list environment (quote, quotation, theorem, definition, enumerate, itemize...), the line after the list may have some extra indentation. If this is the case, add \parshape=0 to the end of the list environment.
	\dropcap{T}he absorption of visible or ultraviolet photons, be it naturally or through artificial light fields, brings molecules into an electronically excited state. Through a variety of mechanisms, the excess energy is either re-emitted though fluorescence, dissipated to some environment, or processed in a photochemical reaction. Ultrafast and radiationless relaxation mechanisms back to the electronic ground state are enabled by \acp{coin} \cite{Yarkony1996,Domcke2011a}. These are ubiquitous regions on the excited state nuclear landscape of molecules, where the energies of two or more states become degenerate. The electronic and nuclear degrees of freedom are then strongly coupled, leading to the breakdown of the adiabatic Born-Oppenheimer picture. Upon optical excitation, an evolving nuclear \ac{wp} can cross into a lower electronic state, if \acp{coin} are within reach. Prominent examples where this process principally determines the outcome of a photochemical process include the primary event of vision \cite{Polli2010}, and the ability of nucleobases as the building blocks of the genetic code to repair the damage of UV radiation \cite{Schultz2004}. 
	
	The existence of \acp{coin} is widely accepted based on simulations, yet has so far eluded direct experimental observation. Indirect experimental evidence for \acp{coin} is usually based on the observation of ultrafast internal conversion rates via electronic state population dynamics \cite{Polli2010,McFarland2014,Galbraith2017}. Different approaches for visualizing \acp{coin} introduced recently include Coulomb Explosion Imaging \cite{Corrales2019}, multidimensional approaches \cite{Oliver2015,Wu2019} or Ultrafast Electron Diffraction \cite{Yang2018}. The decisive impact of \acp{coin} on textbook examples of ultrafast dynamics, like the ring-opening of 1,3--cyclohexadiene \cite{Wolf2019}, has been revealed with unprecedented precision. A novel access to \ac{coin} detection has been enabled recently by the emergence of coherent XUV and x--ray sources \cite{Kowalewski2015,Neville2016,Attar2017,Neville2018,VonConta2018,Kobayashi2019,Timmers2019}. Their unique temporal and spectral profiles can access faster timescales and wider energy windows for spectroscopic measurements \cite{Young2018}.
	
	Here, we show how free electron lasers can be used to monitor \ac{coin} dynamics of the RNA-nucleobase uracil in great detail. A hybrid combination of phase-controlled attosecond (as) and fs x--ray pulses is used to probe the non-adiabatic passage via a stimulated Raman process. A vibronic coherence is generated at the \ac{coin}, which shows up in the signal and its phase at specific Raman shifts, and offers a novel window into the quantum dynamics at these points. The recorded signal reveals the time-resolved distribution of energy splittings of the vibronic states, which provides unique fingerprints of the \ac{coin} pathway. The resulting ultrafast vibronic coherence maps reflect the exact electron/nuclear eigenstates and are independent of any basis set. Spatially localized nuclear \acp{wp} offer a sensitive probe for the energetic topology variation across the \ac{coin}. This spectroscopic access to the \ac{wp} coherence at the \ac{coin} reveals surprising features, with implications for the observation and control of \ac{coin} processes.
	
	\section*{X--Ray TRUECARS Signal of the Uracil Conical Intersection}

To spectroscopically monitor the uracil \ac{coin} passage, we use the TRUECARS technique \cite{Kowalewski2015} that employs a stimulated Raman process induced by a hybrid narrowband and a broadband x--ray pulse. Originally introduced for an ideal model system \cite{Kowalewski2015}, here we use a realistic \textit{ab-initio} molecular Hamiltonian, which we describe in the following.

\begin{figure}[!t]
	\noindent\begin{centering}
		\includegraphics[width=0.98\linewidth]{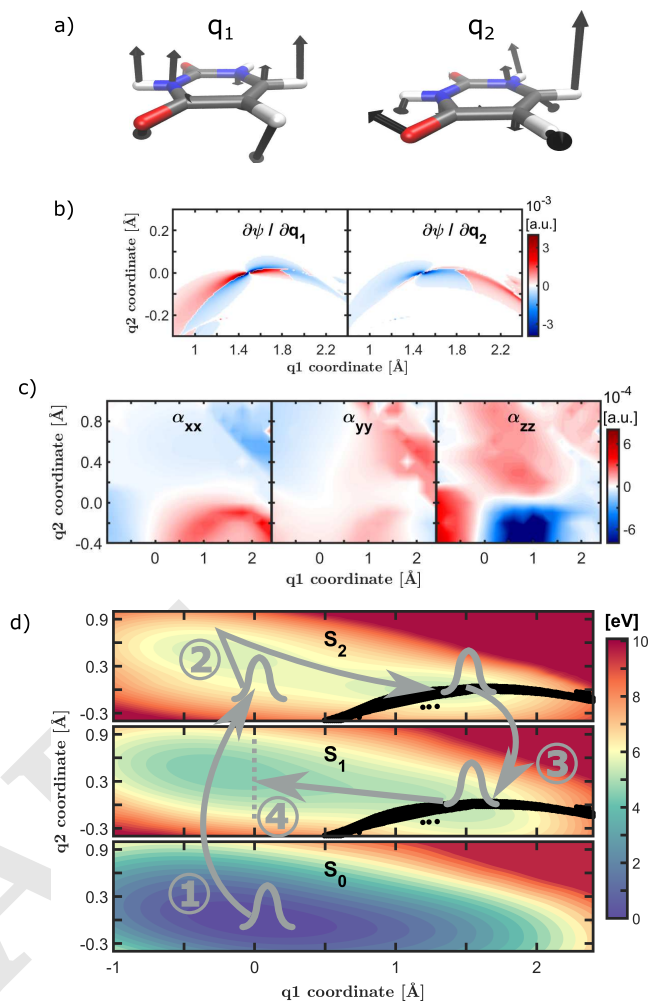}
		\par\end{centering}
	\caption{Wavepacket dynamics of uracil and key molecular quantities for the signal calculation. \textbf{a)} Uracil structure at the ground state minimum geometry, with the vectors indicating the molecular displacement according to the two nuclear degrees of freedom. \textbf{b)} Non-adiabatic couplings between \stwo and \sone at the \ac{coin} seam versus the two nuclear coordinates of the molecules. The characteristic peaked switch of the sign is cleary visible. \textbf{c)} \stwo -- \sone transition polarizabilities (Eq.~(S7)) at 354~eV. The change of electronic character at the conical intersection, where the non-adiabatic couplings have significant values (compare b) ), is clearly visible in the bottom right corners. \textbf{d)} Potential energy surfaces of the three electronic states with the ground state \szero, the dark $n\pi^{\ast}$ \sone state, and the bright $\pi\pi^{\ast}$ state. The conical intersection seam is marked in black, and the wavepacket path is sketched in gray. 1: Optical excitation from \szero to \stwo using a 34~fs UV pump. 2: Free evolution in the \stwo state through the local minimum, with subsequent barrier crossing toward the conical intersection. 3: Non--adiabatic passage through the conical intersection, creating a vibronic coherence. 4: Relaxation in the \sone state toward the energetic minimum, where the wavepacket is absorbed.}
	\label{fig:qc}
\end{figure}

The two nuclear degrees of freedom that span the relevant coordinate space (Fig.~\ref{fig:qc}a)) have been constructed from structures that were found to be important in uracil \stwo photorelaxation \cite{Keefer2017,Matsika2013}, and have originally been described in \cite{Keefer2017}. The first coordinate $q_1$ describes the motion from the \ac{fc} to the \ac{coin}, and is mainly an out--of--plain motion of the initially planar ring structure. The second coordinate $q_2$ represents the motion from the \ac{fc} to a local energetic minimum in the \stwo state, and describes the bending of one Hydrogen atom out of the ring plane. The excited state \acp{pes} in this 2D nuclear space are displayed in  Fig.~\ref{fig:qc}d), and have been shown to reproduce experimental kinetic rates following optical excitation \cite{Keefer2017,Matsika2013}. This reduced-dimensional space has been verified by higher-dimensional treatments \cite{Keefer2017}. The \acp{nac} that are necessary to simulate the passage of the nuclear \ac{wp} through the \ac{coin} are displayed in Fig~ \ref{fig:qc}b). They exhibit a smooth variation with nuclear coordinates, with the characteristic sign change due to the switching of electronic character of the adiabatic states. With the \acp{pes} and the \acp{nac}, nuclear \ac{wp} simulations were launched by exactly solving the \ac{tdse} for our effective Hamiltonian. The full protocol is given in the method section. The nuclear dynamics is launched using a 34~fs \ac{fwhm} Gaussian laser pump in resonance with the \szero $\rightarrow$ \stwo transition, as was done experimentally \cite{Matsika2013}. The path of the \ac{wp} is drawn in Fig.~\ref{fig:qc}d). After excitation to the \stwo state, and following a free \stwo evolution period, the \ac{wp} reaches the \ac{coin} around 100~fs. Here, a vibronic coherence between the \stwo and \sone electronic state is created due to the nuclear \ac{wp} relaxing through the \ac{coin}. Both electronic states are now populated, and the overlap integral of the \stwo and the \sone nuclear \ac{wp} becomes finite. This coherence is probed by the \tc signal. The \ac{wp} parts that reach the \sone state evolve away from the \ac{coin} toward the energetic minimum. There, it is absorbed by using a Butterworth filter \cite{Butterworth1930}, preventing the \ac{wp} from oscillating through the \sone minimum and evolving back to the \ac{coin} region. Our coordinate space is tailored to describe the \stwo evolution, and does not accurately capture \ac{wp} evolution in the \sone far from the \ac{coin}, as other nuclear degrees of freedom start to become important. For example the exit-channel for the \ac{wp} to further \acp{coin} with the \szero state is not included. Back-evolution of the \ac{wp} to the \ac{coin} region should therefore lead to artifacts in the coherence, and absorbing the \ac{wp} in the \sone minimum should describe reality more accurately.

\begin{figure}[!h]
	\noindent\begin{centering}
		\includegraphics[width=0.92\linewidth]{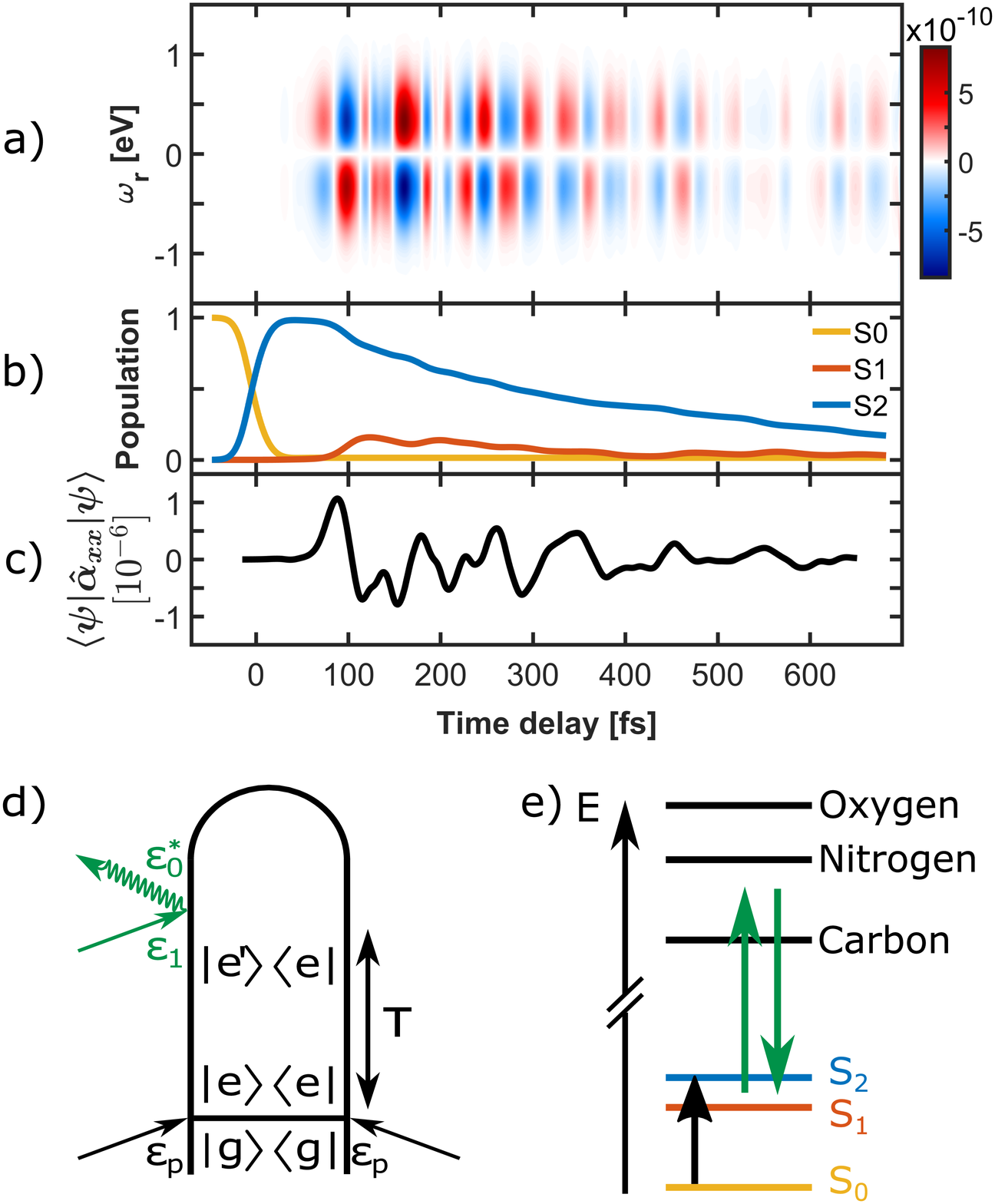}
		\par\end{centering}
	\caption{\textbf{TRUECARS signal of uracil}. \textbf{a)} The frequency-dispersed signal $S(\omega_r,T)$ (Eq.~(\ref{eq:truecars})), using $\hat{\alpha}_{xx}$ at 354~eV (Fig.~\ref{fig:qc}c)). \textbf{b)} Valence-state population dynamics after an optical pump and the dynamics sketched in Fig.~\ref{fig:qc}d). \textbf{c)} Expectation value of the polarizability operator $\hat{\alpha}_{xx}$ (Eq.~\ref{eq:aexpecmat}). $T=0$ is set to the maximum intensity of the 34~fs \ac{fwhm} Gaussian pump. \textbf{d)} Loop diagram for the signal. The Gaussian pump $\varepsilon_p$ creates an electronic and nuclear population in the excited state. A coherence between \textit{e} (\stwo) and \textit{e'} (\sone), that emerges at the conical intersection, is probed by the hybrid field $\varepsilon_0$ (500~as) and $\varepsilon_1$ (2~fs). For diagram rules, see Ref. \cite{Mukamel2010}. \textbf{e)} Energy level scheme of the measurement. Electronic state color code is according to b), where blue and red correspond to the states $i$ and $f$ in Eq.~\ref{eq:polari}. The C, N and O core edges at 291~ev, 405~eV and 536~eV and correspond to the states $c$ in Eq.~\ref{eq:polari}. The green arrows represent the off-resonant Raman process.}
	\label{fig:sig}
\end{figure}

The obtained $\psi (T)$ enters the expression for the frequency-dispersed \tc signal \cite{Kowalewski2015} of the \stwo $\rightarrow$ \sone transition
	\begin{equation}
	\begin{aligned}
	S(\omega_r ,T) = &2 \mathcal{I} \int_{- \infty}^{\infty} dt e^{i \omega_r (t-T)} \varepsilon_0^{\ast} (\omega_r) \varepsilon_1 (t-T)  \\
	&\times \langle \psi (T) \vert \hat \alpha \vert \psi (T) \rangle \quad .
	\label{eq:truecars}
	\end{aligned}
	\end{equation} 
$\varepsilon_0$ and $\varepsilon_1$ are the broadband 2~fs and narrowband 500~as x--ray fields, $\hat \alpha$ is the electronic polarizability operator, $\mathcal{I}$ stands for the imaginary part, and $T$ is the delay between the optical pump and the x--ray probe fields. Applied to the \stwoone intersection, the loop diagram and level scheme of the signal are drawn in Fig.~\ref{fig:sig}. An optical UV pump pulse creates a population in the \stwo state. Following a free evolution period, a coherence between \stwo and \sone (denoted $e$ and $e'$ in Fig.~\ref{fig:sig}d)) is created at the \ac{coin}. This coherence is probed at the instantaneous time \textit{T} by the hybrid x--ray pulse configuration. As sketched in Fig.~\ref{fig:sig}e), they are tuned to be off--resonant with any valence--to--core transition, and their wavelength is at 354~eV between the Carbon and Nitrogen core absorption edges.
	
The time-dependent polarizability in Eq.~(\ref{eq:truecars}) is the relevant material quantity that results from the wavepacket dynamics. In matrix form, it reads
	\begin{equation}
	\langle \psi \vert \hat \alpha \vert \psi \rangle = 
	\langle \left( \begin{array}{c} \psi_2 \\ \psi_1 \end{array} \right) \vert
	\left( \begin{array}{cc} 
	\alpha_{22} & \alpha_{21} \\
	\alpha_{12} & \alpha_{11}
	\end{array} \right) \vert
	\left( \begin{array}{c} \psi_2 \\ \psi_1 \end{array} \right) \rangle .
	\label{eq:aexpecmat}
	\end{equation} 
The diagonal elements $\alpha_{22}$ and $\alpha_{11}$ are the \stwo and \sone electronic state polarizabilities, and the off-diagonal elements are the transition polarizabilities between them. All elements are operators in the nuclear space. When taking the imaginary part in equation (\ref{eq:truecars}), population terms do not contribute to the signal, and $\alpha_{11}$ and $\alpha_{22}$ can be set to zero. This is an important feature of TRUECARS, that makes it a background-free probe of vibronic coherences that in our model only exist in the \ac{coin} vicinity. Since the electronic state populations are larger than the coherence by around three orders of magnitude, the latter would be masked in more conventional transient Raman probing schemes. The hybrid narrowband/broadband x--ray pulse configuration used in \tc provides the necessary temporal and spectral profiles to capture the energy splitting of the \sone \stwo superposition, and the time resolution to resolve the ultrafast coherences. This is not accessible with visible pulses. 

Each element of $\hat \alpha$ is a second-rank tensor with x, y and z Cartesian components in the molecular frame:
\begin{equation}
\hat \alpha = \left( \begin{array}{ccc} 
\alpha_{xx} & \alpha_{yx} & \alpha_{zx} \\ 
\alpha_{xy} & \alpha_{yy} & \alpha_{zy} \\
\alpha_{xz} & \alpha_{yz} & \alpha_{zz} 
\end{array} \right) \quad .
\label{eq:acart}
\end{equation}
Depending on the orientation of the molecule and the pulse configuration, a certain component of Eq.~\ref{eq:acart} is selected. In general, each component of $\hat{\alpha}$ can be calculated according to \cite{Kiefer1982,Olsen1985}:
\begin{equation}
[ \alpha_{xy} ] _{fi} = \sum_c \left\{ \frac{ \langle f \vert \hat \mu_y \vert c \rangle \langle c \vert \hat \mu_x \vert i \rangle}{\omega_{cf} + \omega_0} + \frac{ \langle f \vert \hat \mu_x \vert c \rangle \langle c \vert \hat \mu_y \vert i \rangle}{\omega_{ci} - \omega_0}\right\}
\label{eq:polari}
\end{equation}
Here, $\hat \mu_x$ and $\hat \mu_y$ are the Cartesian $x$ and $y$ components of the dipole operator in the molecular frame. The summation is over all core-hole states $c$, $\omega_{ci}$ is the energy difference between valence state $i$ and the core state $c$, and $\omega_0$~=~354~eV is the carrier frequency of the probe field taken between the carbon and nitrogen K-edges (see Fig. \ref{fig:qc}). Considering an off-resonant x--ray core process between valences states $i$ and $f$, the summation in eq. \ref{eq:polari} should be over all core-hole states $c$. The key quantities in this equation are the transition dipole moments $\langle f \vert \hat \mu_y \vert r \rangle$ between valence and core states, where $\hat \mu_y$ is the $y$-component of the dipole operator. In this work, we calculated all Cartesian components of the dipole moments between the valence \sone and \stwo states on the multireference level of quantum chemistry (see methods) by including 80 core-hole states. Uracil has 8 non-Hydrogen atoms, and we included the first 10 core-hole states of each atom. Dipole moments were calculated at the 2D nuclear coordinate space on 270 grid points, with subsequent interpolation of the surfaces to the grid for \ac{wp} simulations. Exemplary \stwoone transition polarizability surfaces at $\omega_0 = 354~eV$ are drawn in Fig.~\ref{fig:qc}c). They show a pronounced structure along the nuclear coordinates, and the change of electronic character is clearly visible at the \ac{coin} seam.

The \tc signal of the \stwo $\rightarrow$ \sone transition displayed in Fig.~\ref{fig:sig} exhibits some striking features. At the \ac{coin}, a vibronic coherence is created and time-dependent material quantity, i.e. the polarizability expectation value according to Eq.~\ref{eq:aexpecmat}, becomes finite as well. A signal shows up and remains visible for several hundred~fs, until the vibronic coherence decays. This long-lived vibronic coherence is due to the delocalized nature of the quantum nuclear \ac{wp} on the \stwo surface. Tails of the \ac{wp} continue to reach the \ac{coin}, and even with an instant exit-channel for the \ac{wp} in the \sone state, the coherence survives for several hundred~fs. This has profound implications for the observation and control of quantum events at \acp{coin}. The vibronic coherence, that determines the outcome of the photophysical process, is long-lived and its monitoring will require longer temporal windows rather than precise timings for molecules like uracil. Note that Fig.~\ref{fig:qc} shows the signal with the $xx$ component of the polarizability, which would require oriented molecules. However, the signal can be measured for other orientations as well.
\begin{figure}[!h]
	\noindent\begin{centering}
		\includegraphics[width=0.95\linewidth]{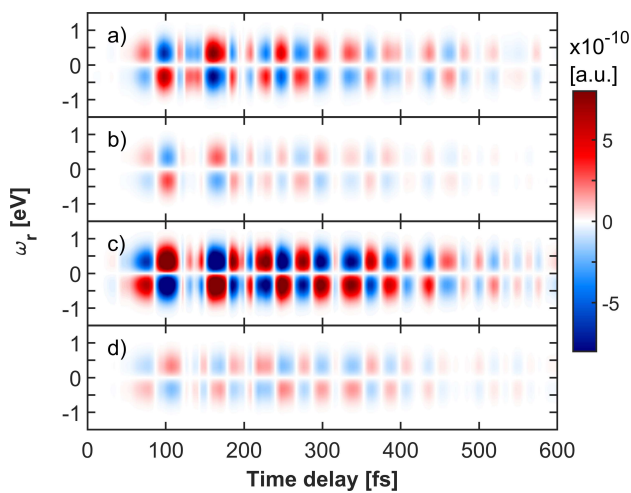}
		\par\end{centering}
	\caption{\textbf{Comparison of molecular orientations in the TRUECARS signal}. \textbf{a)--c)} Signal using $xx$, $yy$ and $zz$ component of the polarizability operator in Eq.~\ref{eq:acart}. The $zz$--component, which is perpendicular to the molecular plane (Fig.~\ref{fig:qc}a)) exhibits the strongest signal, according to the magnitude of the polarizability surfaces in Fig.~\ref{fig:qc}c). \textbf{d)} Isotropic signal, corresponding to a randomly oriented molecular sample. The signal is visible due to the dominating $zz$--contribution.}
	\label{fig:sigxyz}
\end{figure}
This is demonstrated in Fig.~\ref{fig:sigxyz}, where the \tc signal is shown using the $\alpha_{xx}$, $\alpha_{yy}$ and $\alpha_{zz}$ in Fig.~\ref{fig:qc}c). The signal is visible for all three possible molecular orientations. This is even the case for the isotropic signal in Fig.~\ref{fig:sigxyz}d), corresponding to a randomly oriented sample.

\section*{The Time-Resolved Vibronic Frequency Map across the Conical Intersection}
\begin{figure*}[!htb]
	\centering
	\includegraphics[width=0.98\linewidth]{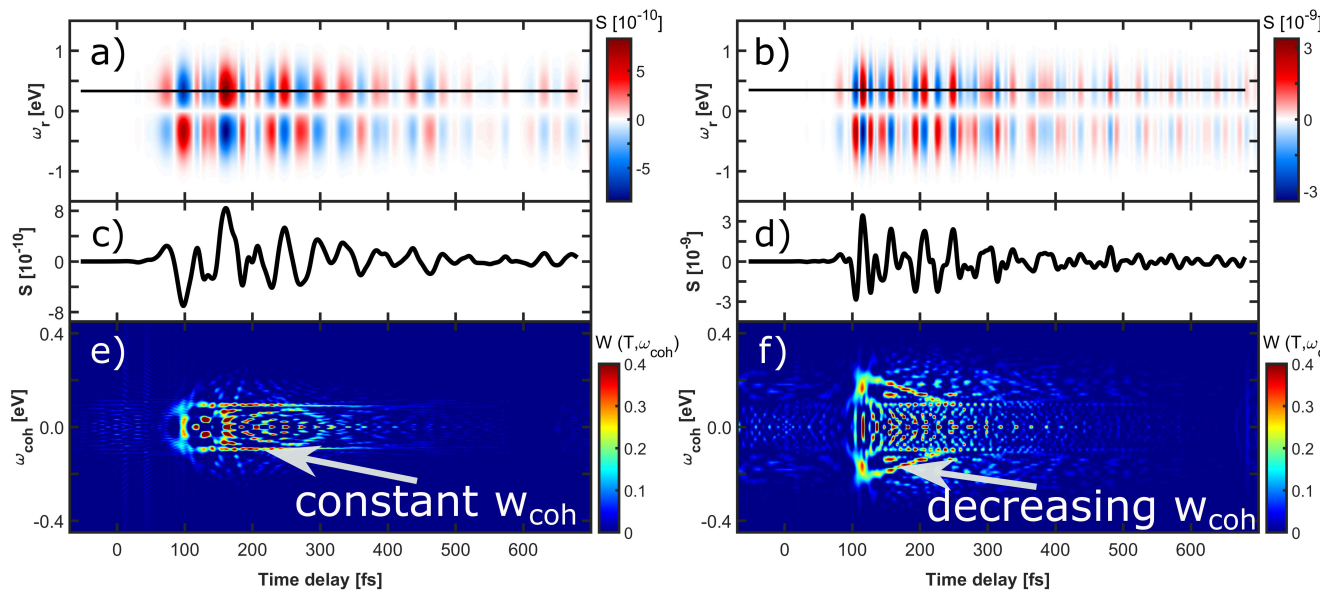}
	\caption{\textbf{Wigner spectrogram (Eq.~\ref{eq:wig}) of the TRUECARS signal}. The left column uses a 34~fs Gaussian pump (delocalized wavepacket). \textbf{a)}: The signal $S(\omega_r,T)$ (Eq.~(\ref{eq:truecars})). \textbf{c)}: Signal trace $S(T)$ at $\omega_r$~=~0.33~eV (maximum signal intensity). \textbf{e)}: Wigner spectrogram $W(T,\omega_{coh})$ of the signal trace $S(T)$ (Eq.~(\ref{eq:wig})). The right column is for a 20~fs Gaussian pump which produces a more localized wavepacket.}
	\label{fig:ttrace}
\end{figure*}

The signal depicted in Fig.~\ref{fig:sig}a) exhibits temporal Stokes and Anti-Stokes oscillations between positive (red) and negative (blue) values. This is because the contributing vibronic states have different frequencies, and the dynamical phase evolution is contained in the signal. To display the time dependent frequency map, we show in Fig.~\ref{fig:ttrace} the Wigner spectrogram of a temporal trace $S(T)$ of the signal (Fig.~\ref{fig:ttrace}b) at a fixed Raman shift $\omega_r$:
\begin{equation}
	W(T,\omega_{coh}) = \int_{- \infty}^{\infty} d \tau S(T+\tau) S(T-\tau) e^{i \omega_{coh} \tau} ,
	\label{eq:wig}
\end{equation}
The~0~to~0.2~eV frequency window spanned by $W(T,\omega_{coh})$ represents the potential energy splitting of the \ac{coin} where the vibronic coherence is located. The main 0.1~eV feature (marked with the white arrow in Fig.~\ref{fig:ttrace}e)) exhibits no temporal variation of $\omega_{coh}$. The contributions visible in the Wigner spectrogram map the vibronic frequencies that contribute to the coherence. Unlike in classical simulations, the complete set of eigenstates is time--independent in our effective Hamiltonian. At all times, the quantum nuclear \ac{wp} can be expressed as a superposition of the vibronic eigenstates, where usually many them contribute. The Wigner spectrogram reveals the dominating contributions to the coherence. Stated differently: When the nuclei are treated as classical, the system has a well-defined, time--dependent energy splitting which vanishes at the conical intersection. In the quantum treatment, this is replaced by a distribution of energy splittings given by the wavepacket frequencies. The average of that distribution serves as the effective splitting.

\begin{figure*}[!htb]
	\centering
	\includegraphics[width=0.98\linewidth]{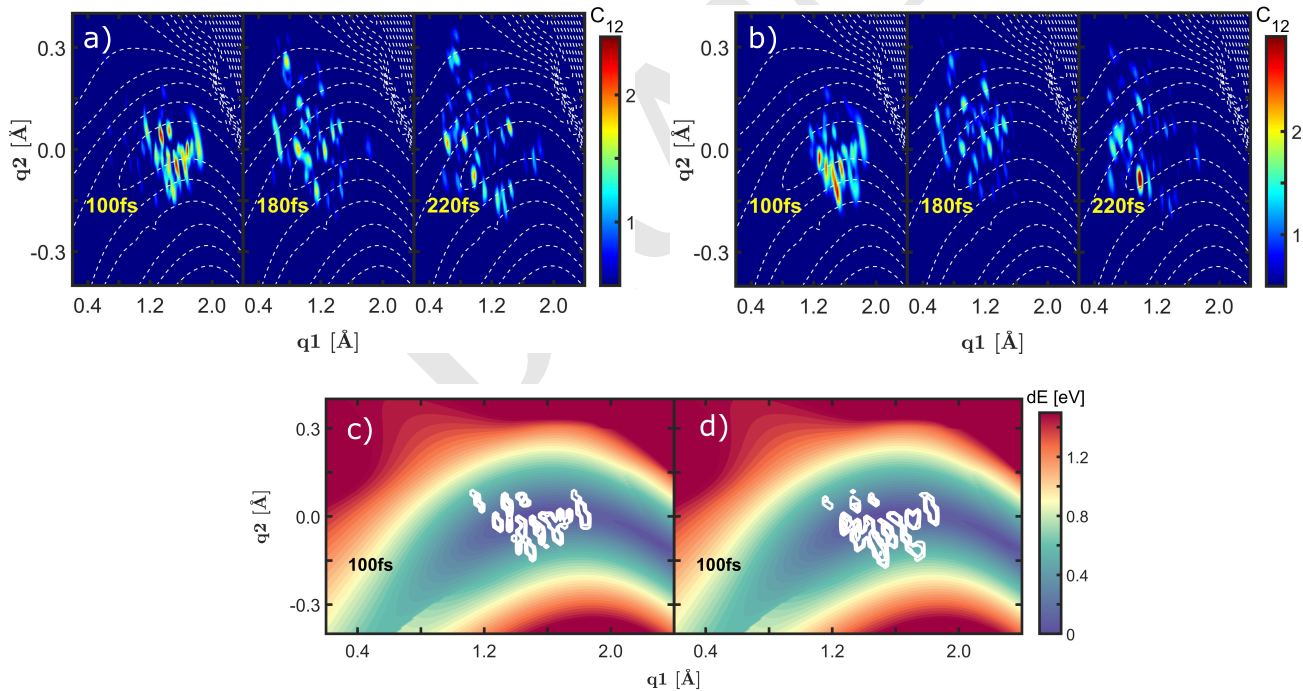}
	\caption{\textbf{Vibronic coherence distribution along the conical intersection path}. Similar to Fig.~\ref{fig:ttrace}, a 34~fs Gaussian pump was used for the left column, and a 20~fs for the right column. \textbf{a)~--~b)}: Vibronic coherence that makes the signal, shown as the normalized product between the \stwo and \sone nuclear wavepacket at 100~fs, 180~fs and 220~fs. The dashed isoline marks the energy difference surface between the adiabatic \stwo and \sone states, with a spacing of 0.2~eV, and the conical intersection seam around $q_2 = 0$. Using the 20~fs pump, the vibronic coherence is more localized at 100~fs and 220~fs, and explores a larger energy splitting. \textbf{c)~--d)}: Different visualization of the 100~fs snapshots in a) and b) with the energy difference E(S\textsubscript{2}) - E(S\textsubscript{1}) as the contour plot. The product of the \stwo and \sone nuclear wavepacket using both pulses is shown at 100~fs (compare a) and b)). In d), the coherence is located at higher splittings than in c). This is directly seen in the corresponding Wigner spectrogram.}
	\label{fig:wigwp}
\end{figure*}

To extend this analysis even further, we had reduced the a 34~fs \ac{fwhm} optical Gaussian pump to 20~fs, as shown in the right column of Fig.~\ref{fig:ttrace}. The \ac{wp} is now more localized in the \stwo \ac{pes}, which strongly influences the coherence. While there is no change in the Raman shift $\omega_{r}$ of the signal, the temporal oscillations are now much faster, so that the frequency regime covered by the vibronic coherence is broader in Fig.~\ref{fig:ttrace}f). More importantly, the energy position of the main feature (marked with the white arrow) does vary with time. Starting at around 0.2~eV at 100~fs, the coherence evolves to lower frequencies until around 250~fs, where it eventually decays. This spectrogram directly maps the distribution of vibronic excitations along the \ac{coin} path, as revealed by the \ac{wp} coherence, and illustrated in Fig.~\ref{fig:wigwp}. Representative snapshots of the normalized \ac{wp} coherence at 100~fs, 180~fs and 220~fs are shown in Fig.~\ref{fig:wigwp}. For the 34~fs pump, the coherence is indeed initially mainly located at energy splittings between 0~eV and 0.2~eV, and delocalizes with time. The position and spread of the coherence in the nuclear phase space is determined by the \acp{wp} in \stwo and \sone. A delocalized coherence does not mean it is lost or has dispersed, it is simply spread over a larger region on the \ac{pes}. In this context, decoherence means that the coherence magnitude, integrated over the nuclear degrees of freedom, decays. This is independent of whether the coherence is local (e.g. at the \ac{coin}), or spread over a larger region.

Using the 20~fs pump (Fig.~\ref{fig:ttrace}b), d) and f), the main  initial coherence at 100~fs is at higher frequencies. After a short delocalization period, the coherence again becomes localized at 220~fs, and shows lower frequencies than at 100~fs. This again corresponds to the position of the vibronic coherence in the nuclear space, depicted in Fig.~\ref{fig:wigwp}. In Fig.~\ref{fig:wigwp}a) and b), the \ac{wp} overlap is drawn as the colored contour plot, while the \stwo \sone energy difference is represented by the isolines. This shows the more local coherence of the 20~fs pump \ac{wp} in contrast to the 34~fs one at 100~fs and 220~fs propagation time. To better identify the range of energy splitting which is spanned by the coherence, and that this corresponds to the main feature in the Wigner spectrograms in Figs.~\ref{fig:ttrace}e) and f), the plotting scheme is inverted in Fig.~\ref{fig:wigwp}c) and d). Here, the \ac{wp} overlap is drawn in isolines, while the \stwo \sone energy splitting is shown with a colored contour plot. The snapshots at 100~fs are shown for both probing schemes, and it can be seen that the \ac{wp} overlap is located at higher energy splittings in Fig.~\ref{fig:wigwp}d) (20~fs pump) than in Fig.~\ref{fig:wigwp}c) (34~fs pump). The transient pathway of the vibronic \ac{wp} coherence is directly accessible from the recorded signal. While we used an adiabatic basis here, the \ac{wp} simulations in the 2D nuclear space of our 2D model Hamiltonian are exact, and therefore independent of the electronic basis. Thanks to the quantum treatment of the nuclei, the time-resolved frequency map in the Wigner spectrogram is an experimental observable, independent of the basis set used, and reveals the exact vibronic state landscape across the \ac{coin} path.
	
\section*{Summary}  
Our study offers a novel window into ultrafast non-adiabatic transitions at \acp{coin}. The frequency, phase and duration of the vibronic coherence, created by the quantum nuclear \ac{wp} evolution, is monitored in real time. Contrary to the common picture of precisely timed \ac{coin} events, the coherence remains visible for several hundred~fs. This is due to the delocalized nature of the quantum nuclear \ac{wp} in the \stwo state after photoexcitation, where small tails consecutively reach the \ac{coin}. The coherence that is built decays due to the \sone \ac{wp} evolving away from the \ac{coin} toward the minimum, but it is constantly rebuilt by other parts arriving at the \ac{coin} in the \stwo state. Localized nuclear \acp{wp} provide a sensitive scan of the frequency landscape surrounding the \ac{coin}. The magnitude and range of the energy splitting of the participating adiabatic vibronic states can be read directly from the signal. No further theoretical input is necessary to access this information from the signal. Uracil photorelaxation is experimentally accessible, and all quantities are produced at the fully \textit{ab-initio} level. The presented Wigner spectrograms offer a novel physical picture of \ac{coin} processes by picturing them through the complex coherence distributions rather than a change in absorption lines or a single vanishing energetic gap. 
	
The off-resonant probing scheme employed here with respect to core excitations renders the signal ubiquitous and translatable to other molecules with a non-vanishing dipole transition between a \ac{coin} valence state and other core-hole states. Monitoring the vibronic coherence provides a unique window how the outcome of ultrafast photophysical processes are decided. This insight is also beneficial for developing approaches for controlling photochemical reactions by modulating this coherence \cite{VondenHoff2012,Arnold2018,Richter2019}. An important step towards phase-controlled XUV/x--ray pulses, that are necessary to record the stimulated Raman signal, was made very recently \cite{Wituschek2020}, and our proposed experiment lies within the capabilities of state of the art x--ray sources \cite{Marinelli2015,Pellegrini2016}. Altogether, the reported novel spectroscopic maps of \ac{coin} passages provide direct access to the quantum \ac{wp} pathways in \ac{coin} passages, which opens the door for a deeper understanding and better control of these ubiquitous processes.

\matmethods{
	\subsection*{Quantum Chemistry}
	The computation of the \acp{pes} and dipole moments for the adiabatic states was described in Ref. \cite{Keefer2017}. This was done at the CASSCF/MRCIS level, using an active space of 12 electrons in 9 orbitals (5 $\pi$, one Oxygen lone pair, and 3 $\pi^{\ast}$ orbitals). For the present study, we had recalculated the \acp{nac} taking into account the phase of the electronic wavefunction. This was done at the CASSCF level and the same (12/9) active space, using the program package MOLPRO \cite{MOLPRO_brief}. A wavefunction file was created containing the selected active space. Using a fine spacing around the \ac{coin}, this wavefunction file was read in for all other \ac{nac} calculations, to guarantee that the sign of the electronic wavefunction remains the same in all structures. The resulting \acp{nac} are shown in Fig.~\ref{fig:qc}b). 
	
	To compute the polarizabilities in Eq.~\ref{eq:polari}, the same active space and level of theory was used with the 6-311G* basis set \cite{Krishnan1980,Clark1983} to calculate 80 core-hole states. Uracil has 8 non-Hydrogen atoms, and we included the first 10 core-hole states of each atom. Dipole moments between the valence \sone and \stwo states to these core states were calculated at the 2D nuclear coordinate space on 270 grid points, with subsequent interpolation of the surfaces to the 256x256 grid for \ac{wp} simulations. Within one calculation, first the valence states were calculated using the (12/9) active space. Then, the lowest energy orbital in the active space was rotated for an s-Orbital of one of the C, N or O cores. This orbital was then frozen, single occupation in this orbital was enforced, and the first 10 excited states were calculated. Afterwards, the spatial components of the dipole operator were calculated between the two valence and 10 core states. This results in a total data set of $270*8*10*2*3=129600$ dipole moments.
	
	\subsection*{Wavepacket Simulations}
	To model the photophysics of uracil, nuclear \ac{wp} simulations were performed by solving the \ac{tdse}
	\begin{equation}
	\begin{aligned}
	i \hbar \frac{\partial}{\partial t} \psi =& \hat H \psi \\
	=& \left[ \hat{T}_q + \hat{V} - \hat \mu \varepsilon (t) \right] \psi	\quad ,
	\end{aligned}
	\end{equation}
	with the kinetic energy operator $\hat{T}_q$ of the nuclei in internal coordinates $q$, the potential energy operator $\hat V$, and the light-matter interaction in the dipole approximation with the dipole operator $\hat \mu$ and the electric field $\varepsilon (t)$. Our model Hamiltonian space has two nuclear degrees of freedom and three electronic states. The \ac{wp} simulation setup has been described and used in Ref. \cite{Keefer2017}.
	
	All quantities are represented on a two-dimensional 256x256 spatial grid. The two nuclear coordinates $q_1$ and $q_2$ are selected from molecular structures that are known to play an important role in uracil \stwo photorelaxation: The \ac{fc}, a local \stwo minimum, and the optimized conical intersection to the \sone state. $q_1$ is the normalized displacement vector that points from the \ac{fc} to the \ac{coin}. $q_2$ is the displacement vector from \ac{fc} to the local \stwo minimum, which is then orthonormalized to \sone. \cite{Keefer2017}
	Numerical propagation using the adiabatic \ac{pes} basis and a nuclear grid is performed by integrating the \ac{tdse} according to
	\begin{equation}
	\psi (t+dt) = exp \left( -i \hat{H} dt  \right) \psi (t) = \hat{U}\psi(t) \quad .
	\end{equation}
	The evolution operator $\hat U$ is expanded in a Chebyshev series according to Ref. \cite{Tal-Ezer1984}, using a time step of 2~a.u. . The kinetic energy operator in internal coordinates $q$ and $s$ is set up according to the G-Matrix formalism \cite{Berens1981} as described in Ref. \cite{Thallmair2016a}.
	\begin{equation}
	\hat{T}_q \simeq - \frac{\hbar ^2}{2m} \sum_{r=1}^{M} \sum_{s=1}^{M} \frac{\partial}{\partial q_r} \left[ G_{rs} \frac{\partial}{\partial q_s} \right] 
	\end{equation}
	with the G-Matrix computed via its inverse elements
	\begin{equation}
	\left( G^{-1} \right)_{rs} = \sum_{i=1}^{3N} m_i \frac{\partial x_i}{\partial q_r} \frac{\partial x_i}{\partial q_s} \quad ,
	\end{equation}
	i.e. via finite differences by displacement of the Cartesian molecular geometry along the internal coordinates. In case of the uracil coordinates described above, the G-Matrix elements are $G_{q_1 q_1} = 0.000167$~a.u., $G_{q_2 q_2} = 0.000139$~a.u., and the kinetic coupling $G_{q_1 q_2} = -0.00007$~a.u. .
	
	A Butterworth filter \cite{Butterworth1930} was employed in the electronic \sone state, which absorbs the parts of the \ac{wp} that reach the \sone minimum. The filter was of "right-pass" type (absorbing all parts on the lef side of the border), and placed at $q_1 =0.5$~\AA{} with an order of 100 \cite{Butterworth1930}. This was done to create an exit channel for the \ac{wp}, which otherwise would move back to the \ac{coin} region after being relaxed from there
	}
	
	\showmatmethods{} % Display the Materials and Methods section
	
	\acknow{We gratefully acknowledge the support of the U.S. Department of Energy, Office of Science, Office of Basic Energy Sciences under Award DE-FG02-04ER15571, and of the National Science Foundation through Grant No. CHE-1953045.
	S.M was supported by the DOE  grant.
	D.K. gratefully acknowledges support from the Alexander von Humboldt foundation through the Feodor Lynen program, and from the continued support of the Dr. Klaus R\"omer foundation. We thank Bing Gu, Stefano Cavaletto and Jérémy Rouxel for clarifying and helpful technical discussions, and Markus Kowalewski for providing his QDng quantum dynamics code.}
	
	\showacknow{} % Display the acknowledgments section
	
	% Bibliography
	\bibliography{lit}
	
\end{document}